\documentclass[aps,showpacs,twocolumn,prl,Letter,superscriptaddress]{revtex4}
\usepackage{amsmath}
\usepackage{amssymb}
\usepackage{mathtext}
\usepackage{graphicx}

\begin{document}

\title{
Dynamics of Limit Cycle Oscillator Subject to General Noise}
\date{\today}

\author{Denis S.\ Goldobin}
\address{Institute of the Continuous Media Mechanics, UB RAS,
         Perm 614013, Russia}
\address{Department of Mathematics, University of Leicester,
         Leicester LE1 7RH, UK}
\author{Jun-nosuke Teramae}
\affiliation{Brain Science Institute, RIKEN, Wako 351--0198, Japan}
\author{Hiroya Nakao}
\affiliation{Department of Physics, Kyoto University, Kyoto 606--8502, Japan}
\author{G.\ Bard Ermentrout}
\affiliation{Department of Mathematics, University of Pittsburgh,
             Pittsburgh, Pennsylvania 15260, USA}

\begin{abstract}
The phase description is a powerful tool for analyzing noisy limit
cycle oscillators. The method, however, has found only limited
applications so far, because the present theory is applicable only
to the Gaussian noise while noise in the real world often has
non-Gaussian statistics. Here, we provide the phase reduction
method for limit cycle oscillators subject to general, colored and
non-Gaussian, noise including heavy-tailed one. We derive
quantifiers like mean frequency, diffusion constant, and the
Lyapunov exponent to confirm consistency of the results. Applying
our results, we additionally study a resonance between the phase
and noise.
\end{abstract}

\maketitle

Limit cycle oscillators effectively model various sustained
oscillations in many fields of science and technology including
chemical reactions, biology, electric circuits, and
lasers~\cite{Kuramoto-2003,population_synchrony,bacterial_colonies,neuronal_networks}.
The phase reduction method is a powerful analytical tool which
approximates high-dimensional dynamics of limit cycle oscillators
with single phase variable that characterizes timing of
oscillation~\cite{Winfree-1967,Kuramoto-2003}. Since the phase is
neutrally stable, phase perturbations persist in time and result
in various remarkable phenomena where weak action leads to
significant effects, such as those addressed in the theory of
synchronization~\cite{Pikovsky-Rosenblum-Kurths-2003,Zhou_etal-2002}.
While the theory of the phase reduction had been developed for
deterministic oscillators, recent studies successfully extended
the theory to limit cycle oscillators subject to
noise~\cite{Yoshimura-Arai-2008,Teramae-Nakao-Ermentrout-2009,Yoshimura-2010}
and revealed that interplay between nonlinearity and noise results
in fascinating noise-induced phenomena including frequency
modulation and noise-induced
synchronization~\cite{Lyapunov_exponent,Lyapunov_exponent-desynchronization}.

This extended phase reduction method, however, has found limited
applications so far, since the method is applicable only to
Gaussian noise. While the noise in the real world often has
non-Gaussian statistics, few theories have considered nonlinear
systems subject to general non-Gaussian noise, which has forced
people to use the Gaussian approximation. In particular, whether
the phase description is still valid for oscillators subject to
non-Gaussian noise and how quantifiers of the phase dynamics
should be amended remains unknown. In this paper, we develop the
phase reduction method for limit cycle oscillators subject to
general, colored and non-Gaussian noise. By correctly evaluating
the influence of amplitude perturbations up to second order in the
noise strength, we derive the stochastic differential equation of
phase, which allows us to study nonlinear oscillations in the real
world without the Gaussian approximation. To confirm consistency
of the result, we derive closed expressions of quantifiers of the
phase dynamics such as mean frequency, phase diffusion constant,
and the Lyapunov exponent. The only limitation we impose is the
weakness of the noise. Thus, the obtained results are applicable
even when higher order moments of the noise diverge as long as the
second order moment is finite and we confirm this fact
numerically. As an application of the results, we study a limit
cycle oscillator driven by a phase noise with a finite correlation
time and show that amended quantifiers precisely predict resonance
between phase and the noise.

We start with the case of a two-dimensional {\em limit cycle
oscillator} and then extend our results to higher dimensions and
multi-component noise. One can describe the evolution of the
system subject to noise in terms of the phase $\phi$ and the
amplitude deviation $r$ from the limit
cycle~\cite{Yoshimura-2010,auxil};
\begin{eqnarray}
 &&\dot{\phi}=\omega+\sigma f(\phi,r)\eta(t)\,,
\label{eq-1-01}\\
 &&\dot{r}=-\lambda r+\sigma g(\phi,r)\eta(t)\,;
\label{eq-1-02}
\end{eqnarray}
here $\omega$ is the cyclic frequency of unperturbed oscillations;
$\lambda:=-(\omega/2\pi)\ln{\Lambda}$ and $\Lambda$ is the Floquet
multiplier of the cycle, {\it i.e.}, $\lambda$ is the average
amplitude relaxation rate; $\eta(t)$ is a normalized noise;
$\sigma\ll1$ is the noise amplitude; $f(\phi,r)$ and $g(\phi,r)$
are $2\pi$-periodic in $\phi$ and represent sensitivity of the
phase and amplitude, respectively, to noise. The amplitude
deviation is nonuniformly scaled so that Eq.\,(\ref{eq-1-02}) is
not an approximation, but uniformly valid over the basin of
attraction of the limit cycle, as we rigorously show in auxiliary
material~\cite{auxil}.

We use $\sigma$ as an expansion parameter;
 $\phi(t)=\phi_0(t)+\sigma \phi_1(t)+\sigma^2\phi_2(t)+...$,
 $r(t)=\sigma r_1(t)+\sigma^2r_2(t)+...$,
 $f(\phi,r)=f_0(\phi)+f_1(\phi)\,r+...$, and
 $g(\phi,r)=g_0(\phi)+g_1(\phi)\,r+...$\,.
From Eqs.\,(\ref{eq-1-01}) and (\ref{eq-1-02}),
 $\phi_0(t)=\omega t$,
 $\dot{\phi}_1=f_0[\phi_0(t)]\eta(t)$, and
 $\dot{r}_1=-\lambda r_1+g_0[\phi_0(t)]\eta(t)$;
the latter two formulae provide
\begin{equation}
 \phi_1(t)=\int_{-\infty}^t f_0[\phi_0(t_1)]\eta(t_1)dt_1\,,
\label{eq-1-03}
\end{equation}
\begin{equation}
 r_1(t)=\int_0^{+\infty} g_0[\phi_0(t)-\omega\tau]\eta(t-\tau)e^{-\lambda\tau}d\tau\,.
\label{eq-1-04}
\end{equation}
Meanwhile, the expansion of Eq.\,(\ref{eq-1-01}) reads
\[
\dot{\phi}=\omega+\sigma f_0(\phi_0)\eta
 +\sigma^2\big[f_0'(\phi_0)\phi_1\eta+f_1(\phi_0)r_1\eta\big]+O(\sigma^3),
\]
here prime denotes derivative with respect to $\phi$. The
right-hand part of the latter equation except for the term
proportional to $f_1(\phi)$ is merely the expansion of
Eq.\,(\ref{eq-1-01}) with $f(\phi,r)$ replaced by $f(\phi,0)$.
Therefore, we can keep the equation unexpanded with respect to
$\phi$ but add the correction owing to $r_1(t)$;
\[
\dot{\phi}=\omega+\sigma f_0(\phi)\eta(t)
 +\sigma^2f_1(\phi_0)r_1\eta(t)+O(\sigma^3)\,.
\]
$\sigma^2f_1(\phi_0)r_1\eta(t)$ is small in comparison to $\sigma
f_0(\phi)\eta(t)$, but makes an average contribution of the same
order (because $\langle\dot{\phi}_1\rangle=0$). Thus, the
fluctuating part of this term is not principal and may be omitted;
\[
\dot{\phi}\approx\omega+\sigma f_0(\phi)\eta(t)
 +\langle\sigma^2f_1(\phi_0)r_1\eta(t)\rangle\,.
\]
Employing expression (\ref{eq-1-04}) for $r_1$, we obtain
\[
 \langle f_1(\phi_0)r_1\eta(t)\rangle
 =f_1[\phi_0(t)]\int_0^{+\infty}\!g_0[\phi_0(t)-\omega\tau]\,C(\tau)e^{-\lambda\tau}d\tau,
\]
where $C(\tau):=\langle\eta(t)\eta(t-\tau)\rangle$ is the noise
autocorrelation function. Finally, the reduced phase equation up
to the leading contributions reads
\begin{equation}
\dot{\phi}=\omega+\sigma f_0(\phi)\eta(t)
 +\frac{\sigma^2}{\omega}f_1(\phi)\!\!
 \int_0^{+\infty}\!\!g_0(\phi-\psi)\,C\!\left(\frac{\psi}{\omega}\right)e^{-\frac{\lambda}{\omega}\psi}d\psi.
\label{eq-1-05}
\end{equation}
Here $\tau$ is replaced with $\psi/\omega$; the corrections to
$\dot{\phi}$ caused by replacement of $\phi_0$ with $\phi$ in the
integrand are $\propto\sigma^3$ and thus negligible. Remarkably,
the effect of the amplitude relaxation rate $\lambda$ can be
approximately interpreted as cutting-off long-term
auto-correlations of noise if there are some, because for large
$\lambda\tau$ correlation function $C(\tau)$ is suppressed by the
exponential factor. Thus $\lambda^{-1}$ determines the maximal
efficient range of noise auto-correlation.

For Ornstein--Uhlenbeck noise,
$C(\tau)=\gamma\exp(-\gamma|\tau|)$, the reduced phase equation
(\ref{eq-1-05}) takes the form
\[
\dot{\phi}=\omega+\sigma f_0(\phi)\eta(t)
 +\frac{\sigma^2\gamma}{\omega}f_1(\phi)\!\!
 \int_0^{+\infty}\!\!g_0(\phi-\psi)e^{-\frac{\lambda+\gamma}{\omega}\psi}d\psi\,,
\]
which coincides with the one presented in
Ref.\,\cite{Yoshimura-2010} and implies the corresponding results
of
Refs.\,\cite{Yoshimura-Arai-2008,Teramae-Nakao-Ermentrout-2009,Galan-2009}.
While Ref.\,\cite{Teramae-Nakao-Ermentrout-2009} considers the
case of Gaussian noise, a highly stable limit cycle and short
noise correlation times and Ref.\,\cite{Yoshimura-2010} is limited
to the case of OU noise, the present theory includes their results
(as special cases) and additionally allows dealing with
non-Gaussian noise, arbitrary noise auto-correlation functions
(including signals of chaotic oscillators) and arbitrary rate of
amplitude relaxation.

The procedure for deriving the reduced phase equation suggests
that this equation will provide the correct probability density
function for $\phi$ and mean frequency
$\Omega\equiv\langle\dot{\phi}\rangle$ up to $O(\sigma^2)$;
\begin{eqnarray}
 &&\hspace{-10pt}\Omega=\omega+
 \frac{\sigma^2}{\omega}\left\langle
  f_0'(\phi)\!\int_0^{+\infty}\!f_0(\phi-\psi)\,C\!\left(\frac{\psi}{\omega}\right)
  d\psi\right\rangle_{\!\!\phi}
\nonumber\\
 &&+\frac{\sigma^2}{\omega}\left\langle
  f_1(\phi)\!\int_0^{+\infty}\!g_0(\phi-\psi)\,C\!\left(\frac{\psi}{\omega}\right)
  e^{-\frac{\lambda}{\omega}\psi}d\psi\right\rangle_{\!\!\phi}
   \label{eq-1-06}
\end{eqnarray}
[henceforth,
$\langle...\rangle_\phi\equiv(2\pi)^{-1}\int_0^{2\pi}...d\phi$].
The noise can either increase or decrease the mean frequency,
depending on features of correlation function $C(\tau)$,
sensitivity functions, and the cycle stability ({\it e.g.}, see
Fig.\,\ref{fig2}). However, one should verify whether the more
subtle quantities---the phase diffusion constant $D$ and the
leading Lyapunov exponent $\lambda_0$---can be correctly evaluated
from Eq.\,(\ref{eq-1-05}).

The principal contributions to the phase diffusion are readily
determined from Eq.\,(\ref{eq-1-05}); indeed,
\begin{eqnarray}
 &&\hspace{-10pt}D=\int_{-\infty}^{+\infty}\langle(\dot\phi(t)-\langle\dot{\phi}\rangle)
 (\dot\phi(t+\tau)-\langle\dot{\phi}\rangle)\rangle d\tau
\nonumber\\
 &&=\sigma^2\int_{-\infty}^{+\infty}\langle\dot\phi_1(t)\dot\phi_1(t+\tau)\rangle d\tau+O(\sigma^4)
\nonumber\\
 &&\hspace{-10pt}=\frac{\sigma^2}{2\pi}\int_0^{2\pi}\!d\phi
 \int_{-\infty}^{+\infty}\!d\tau\,f_0(\phi)\,f_0(\phi+\omega\tau)\,C(\tau)+O(\sigma^4);
 \label{eq-1-07}
\end{eqnarray}
$\dot\phi_1(t)$ [Eq.\,(\ref{eq-1-03})] is precisely determined by
terms accounted in Eq.\,(\ref{eq-1-05}); therefore,
Eq.\,(\ref{eq-1-07}) is completely consistent with the reduced
phase equation. Interestingly, up to the leading order of accuracy
the phase diffusion is not affected by the extra amplitude terms.
Thus, for instance, the analytical results and important
conclusions of
Refs.~\cite{Goldobin-Rosenblum-Pikovsky-2003,Goldobin-2008} for
limit cycle oscillators subject to weak noise and delayed feedback
control remain correct.

For the leading Lyapunov exponent, the situation is more subtle.
To deal with it rigorously, we consider a small perturbation
$(\alpha=\alpha_0\exp[\mu(t)],s)$ to the solution $(\phi(t),r(t))$
of Eqs.\,(\ref{eq-1-01}) and (\ref{eq-1-02}). We have
\begin{eqnarray}
 &&\hspace{-6pt}
 \dot{\mu}=\sigma(f_0'[\phi(t)]+r(t)f_1'[\phi(t)])\eta(t)
 +\sigma f_1[\phi(t)]\frac{s}{\alpha_0}\eta(t)e^{-\mu}\,,
\nonumber\\
 &&\hspace{-6pt}
 \dot{s}=-\lambda s+\sigma g_0'[\phi(t)]\alpha_0e^\mu\eta(t)
 +\sigma g_1[\phi(t)]s\eta(t)\,.
\nonumber
\end{eqnarray}
and employ the standard multiscale method adopting
 $\mu(t)=\mu(t_0,t_2,...)$,
 $d/dt=\partial/\partial t_0+\sigma^2\partial/\partial t_2+...$, {\it etc}.
After some calculations, one finds the expression for the leading
Lyapunov exponent $\lambda_0:=\langle\dot\mu\rangle$ up to
$O(\sigma^2)$:
\begin{eqnarray}
 &&\hspace{-10pt}\lambda_0=\frac{\sigma^2}{\omega}\Big\langle f_0''(\phi)
 \int_0^{+\infty}f_0(\phi-\psi)\,C\!\left(\frac{\psi}{\omega}\right)\,d\psi
\nonumber\\
 &&+\frac{\partial}{\partial\phi}\Big[f_1(\phi)
 \int_0^{+\infty}g_0(\phi-\psi)\,C\!\left(\frac{\psi}{\omega}\right)
 e^{-\frac{\lambda\psi}{\omega}}\,d\psi\Big]\Big\rangle_\phi
 +O(\sigma^4)
\nonumber\\
 &&=-\frac{\sigma^2}{\omega}\Big\langle
 f_0'(\phi)\int_0^{+\infty}f_0'(\phi-\psi)\,C\!\left(\frac{\psi}{\omega}\right)\,d\psi\Big\rangle_\phi
 +O(\sigma^4),
\label{eq-1-08}
\end{eqnarray}
which is consistent with the phase equation (\ref{eq-1-05}). Note,
in the latter equations, the amplitude degree of freedom, which
was disregarded in previous works, impacts the instantaneous
growth rate of perturbations, but averages out to zero. Thus, on
the one hand, our results demonstrate the importance of amplitude
degrees of freedom for the stability of response of a general
limit cycle oscillator even in the limit of vanishing noise; on
the other hand, its average impact turns out to be zero up to the
leading order of accuracy for general noise, proving that
analytical calculations and conclusions presented
in~\cite{Lyapunov_exponent,Goldobin-2008} are valid for real
situations. Notice, the negative Lyapunov exponent and its
decrease with increase of the noise strength are related to the
stability of the noisy system response in sense that it attracts
trajectories (the phenomenon is known as noise-induced
synchronization), but this does not mean that the response is
regular due to the nonzero phase diffusion.

All the results can be  extended in a straightforward manner to
the case of an $N$-dimensional dynamical system subject to
$M$-component noise;
\begin{eqnarray}
&&\hspace{-10pt}
\dot{\phi}=\omega+\sum_{\beta=1}^M\Bigg[\sigma_\beta f_\beta(\phi,{\bf 0})\,\eta_\beta(t)
 +\sum_{j=1}^{N-1}\frac{\sigma_\beta^2}{\omega}
 \left(\frac{\partial f_\beta(\phi,{\bf r})}{\partial r_j}\right)_{{\bf r}={\bf 0}}
\nonumber\\
&&\qquad
 \times
 \int_0^{+\infty}g_\beta(\phi-\psi,{\bf 0})\,C_\beta\!\left(\frac{\psi}{\omega}\right)e^{-\frac{\lambda_j\psi}{\omega}}d\psi\Bigg],
\label{eq-2-01}
\end{eqnarray}
\begin{equation}
D=\sum_{\beta=1}^M\frac{\sigma_\beta^2}{\omega}
 \Big\langle f_\beta(\phi,{\bf 0})\int_{-\infty}^{+\infty}f_\beta(\phi-\psi,{\bf 0})\,
 C_\beta\!\left(\frac{\psi}{\omega}\right)d\psi\Big\rangle_\phi,
\label{eq-2-02}
\end{equation}
\begin{equation}
\lambda_0=-\sum_{\beta=1}^M\frac{\sigma_\beta^2}{\omega}
 \Big\langle \frac{\partial f_\beta(\phi,{\bf 0})}{\partial\phi}
 \int_0^{+\infty}\frac{\partial f_\beta(\phi-\psi,{\bf 0})}{\partial\phi}\,
 C_\beta\!\left(\frac{\psi}{\omega}\right)d\psi\Big\rangle_\phi.
\label{eq-2-03}
\end{equation}
Here $\beta$ indexes noise components, $j$ does the degrees of
freedom transversal to the limit cycle.

Now, we address the issue of applicability of our results for
noise with diverging higher moments. Although the derived
expressions involve only second moments of the noise, one has to
check that possible divergence of higher moments does not break
the entire expansion and influence $\Omega$, $D$, and $\lambda_0$
in the main order.

\begin{figure}[!t]
\center{\sf
 (a)\hspace{-5.5mm}
 \includegraphics[width=0.222\textwidth]%
 {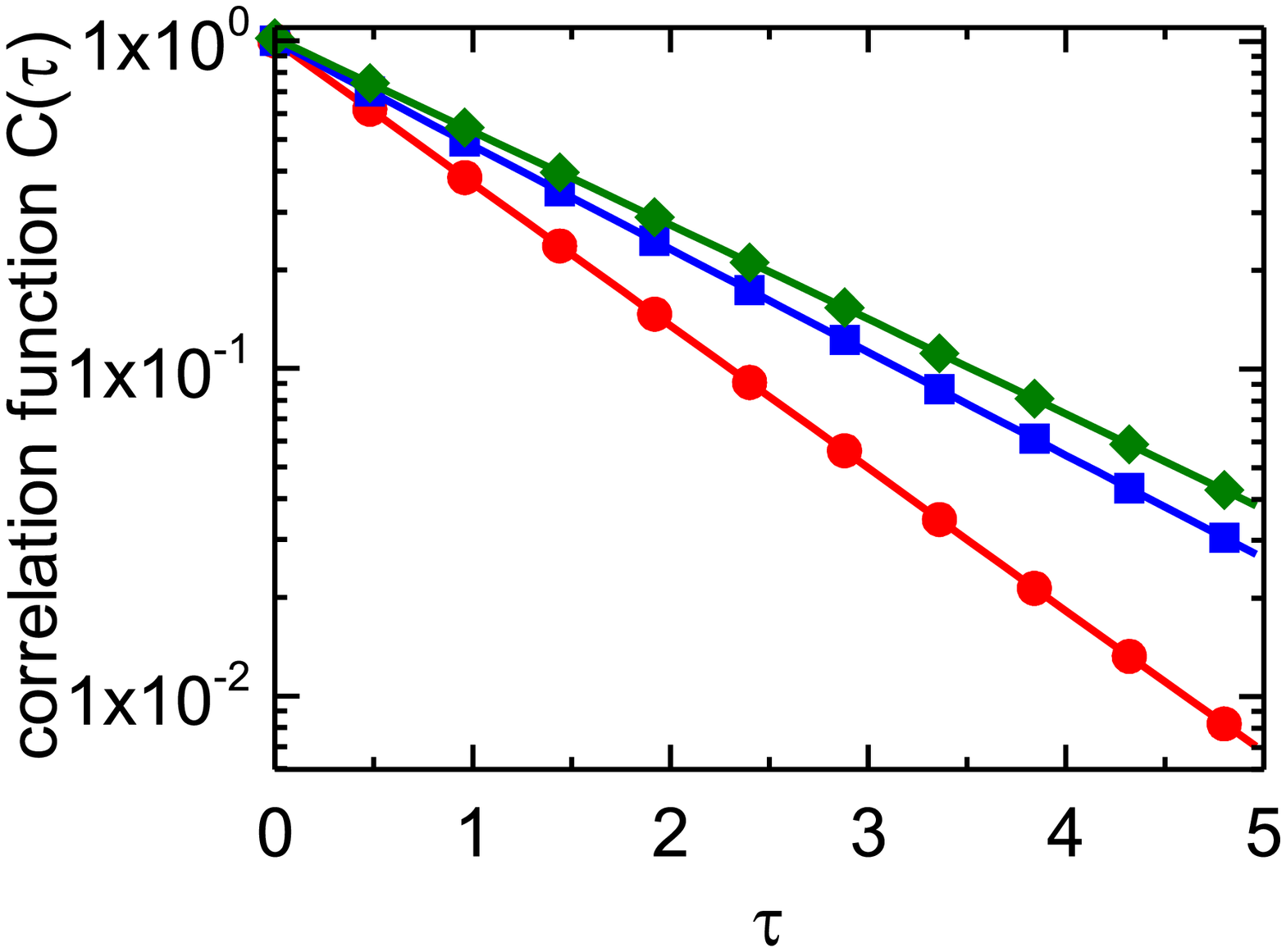}
 \quad
 (b)\hspace{-5.5mm}
 \includegraphics[width=0.222\textwidth]%
 {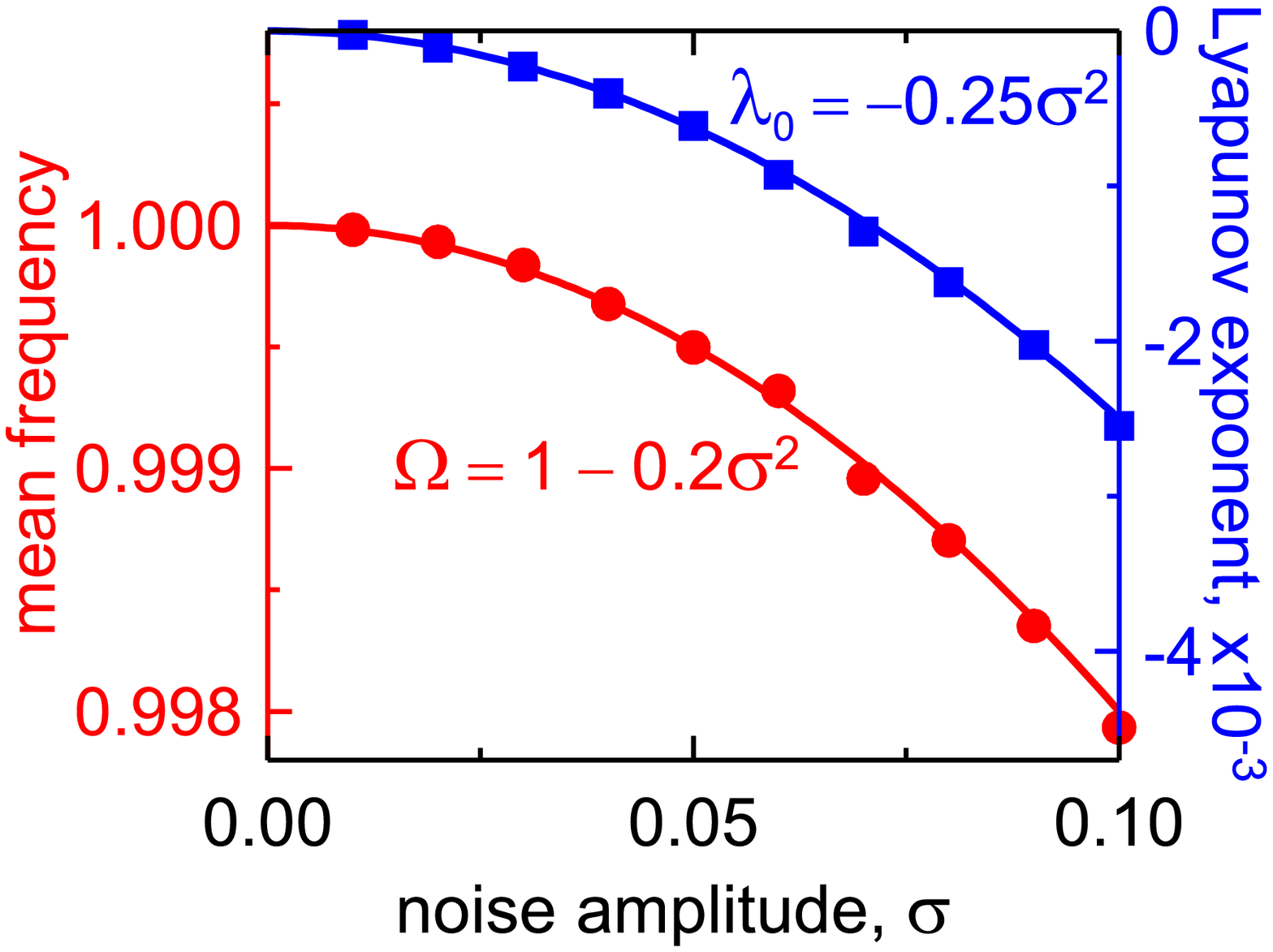}
 \\[4pt]
 (c)\hspace{-5.5mm}
 \includegraphics[width=0.222\textwidth]%
 {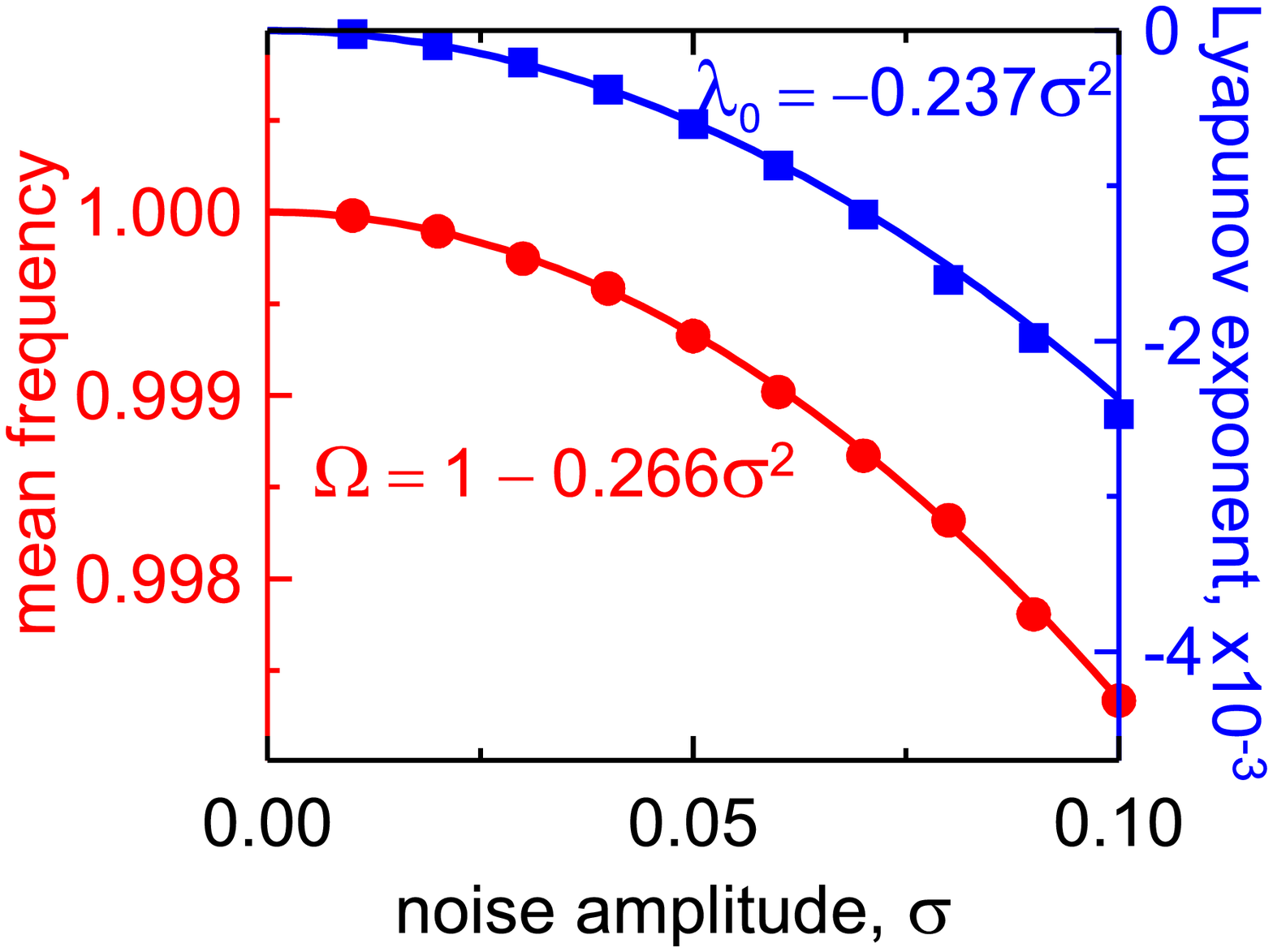}
 \quad
 (d)\hspace{-5.5mm}
 \includegraphics[width=0.222\textwidth]%
 {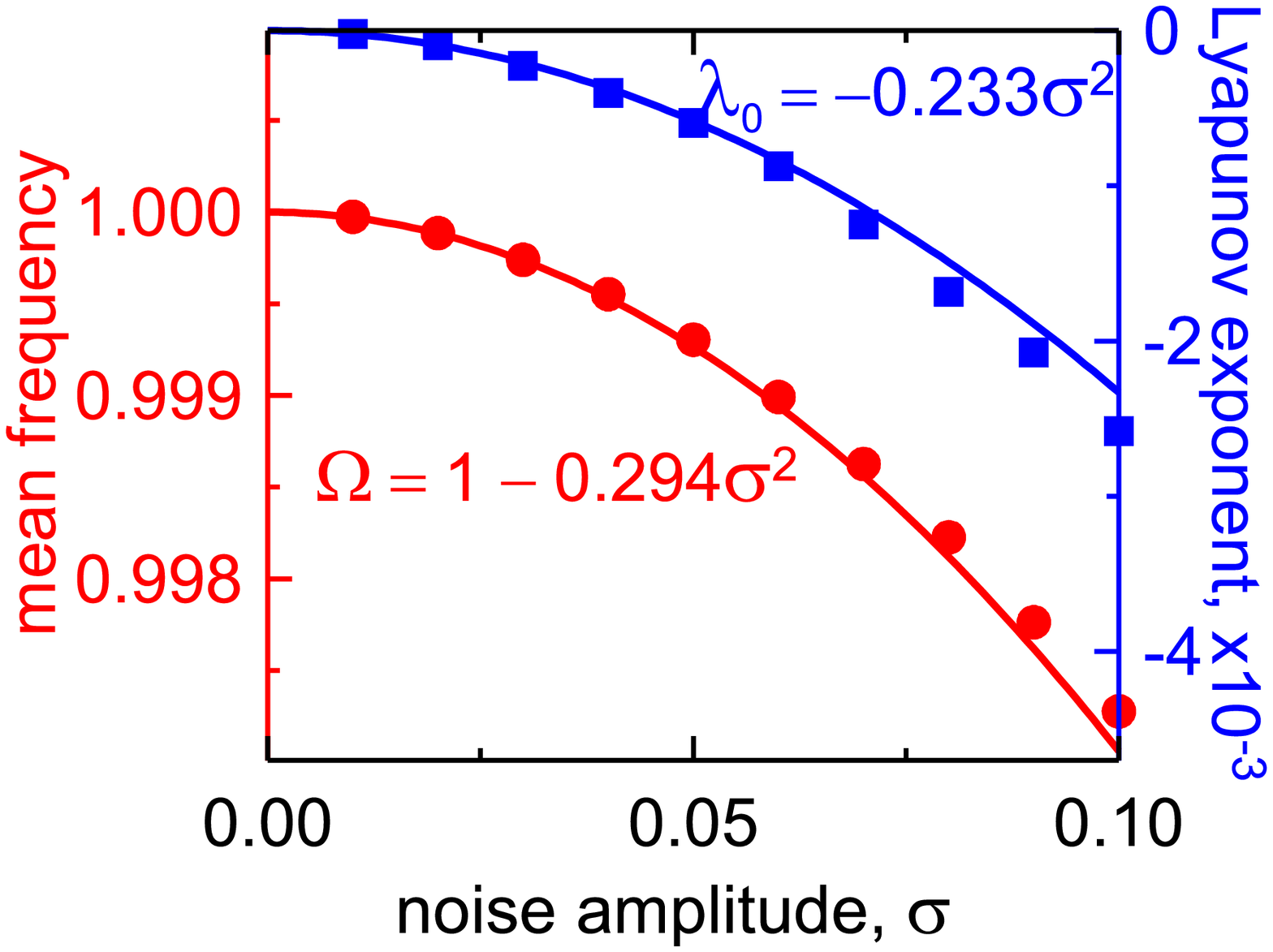}
}
  \caption{(Color online) Hopf oscillator (\ref{eq-3-01})
subject to different noises; here $\tau_\eta=1$ and $\lambda=2$.
\\
(a): correlation function $C(\tau)$ for Ornstein--Uhlenbeck noise,
which is Gaussian, (red circles) and noises with exponential (blue
squares) and fractional rational (green diamonds) distributions.
\\
(b)--(d): the numerically calculated mean frequency (red circles)
and Lyapunov exponent (blue squares) are in good agreement with
Eqs.\,(\ref{eq-3-03}) and (\ref{eq-3-04}) (solid lines) for OU
noise (b) and noises with exponential (c) and fractional rational
(d) distributions.
 }
  \label{fig1}
\end{figure}

For this reason we performed numerical simulation of a Hopf
oscillator subject to colored noise $\eta(t)$:
\begin{eqnarray}
&&\dot{A}=iA+(\lambda/2)(1-|A|^2)A+\sigma\eta\,,
\label{eq-3-01}\\
&&\dot{\eta}=\tau_\eta^{-1}[-\eta+s(\eta)\,\xi(t)]\,,
\label{eq-3-02}
\end{eqnarray}
where $A$ is complex, the noise acts only on $\mathrm{Re}(A)$
(Eq.\,(\ref{eq-3-01}) describes, for instance, lasers with optical
injection in the limit of large density of excited states;
cf.~\cite{Lang-1982}), $\xi(t)$ is Gaussian and white:
$\langle\xi(t)\xi(t')\rangle=2\delta(t-t')$. We consider
normalized noises $\eta(t)$ ($\langle\eta^2\rangle=1$) with three
kinds of distribution $V(\eta)$:
\\
(1)\;Gaussian, $V_1(\eta)=(2\pi)^{-1/2}\exp(-\eta^2/4)$;
\\
(2)\;exponential, $V_2(\eta)=(1/4)\exp(-|\eta|/2)$, which has
nonzero but still finite higher cumulants; and
\\
(3)\;fractional rational function,
$V_3(\eta)=\pi^{-1}(1+\eta^2)^{-2}$, for which
$\langle\eta^{2n}\rangle$ is finite only for $n=1$.
\\
These noises are generated with employment of
 $s_1(\eta)=1$,
 $s_2(\eta)=\sqrt{1/4+|\eta|/2}$, and
 $s_3(\eta)=\sqrt{(1+\eta^2)/3}$ in Eq.\,(\ref{eq-3-02}).
For the oscillator (\ref{eq-3-01}), one finds $f_0=-\sin{\phi}$,
$f_1=-f_0=\sin{\phi}$, and $g_0=\cos{\phi}$; therefore,
\begin{eqnarray}
&&\Omega=1-\frac{\sigma^2}{2}\int_0^\infty\sin{\psi}\,(1-e^{-\lambda\psi})\,C(\psi)\,d\psi\,,
\label{eq-3-03}
\\
&&D=-2\lambda_0=\sigma^2\int_0^\infty\cos{\psi}\,C(\psi)\,d\psi\,.
\label{eq-3-04}
\end{eqnarray}
For exponential and fractional rational distributions, the
correlation function $C(\tau)$ was calculated numerically. In
Fig.\,\ref{fig1} one can see that the analytical theory is in
fairly good agreement with results of numerical simulation both
for noises with all moments finite (b, c) and for one with
infinite $\langle\eta^4\rangle$ (d). For the latter case the
analytical theory is practically no less accurate than for the
former ones though, for strong noise, the mismatch between theory
and numerics is more pronounced because of large fluctuations
occurring in distributions with heavy tails.

\begin{figure}[!t]
\center{\sf
 (a)\hspace{-5mm}
 \includegraphics[width=0.222\textwidth]%
 {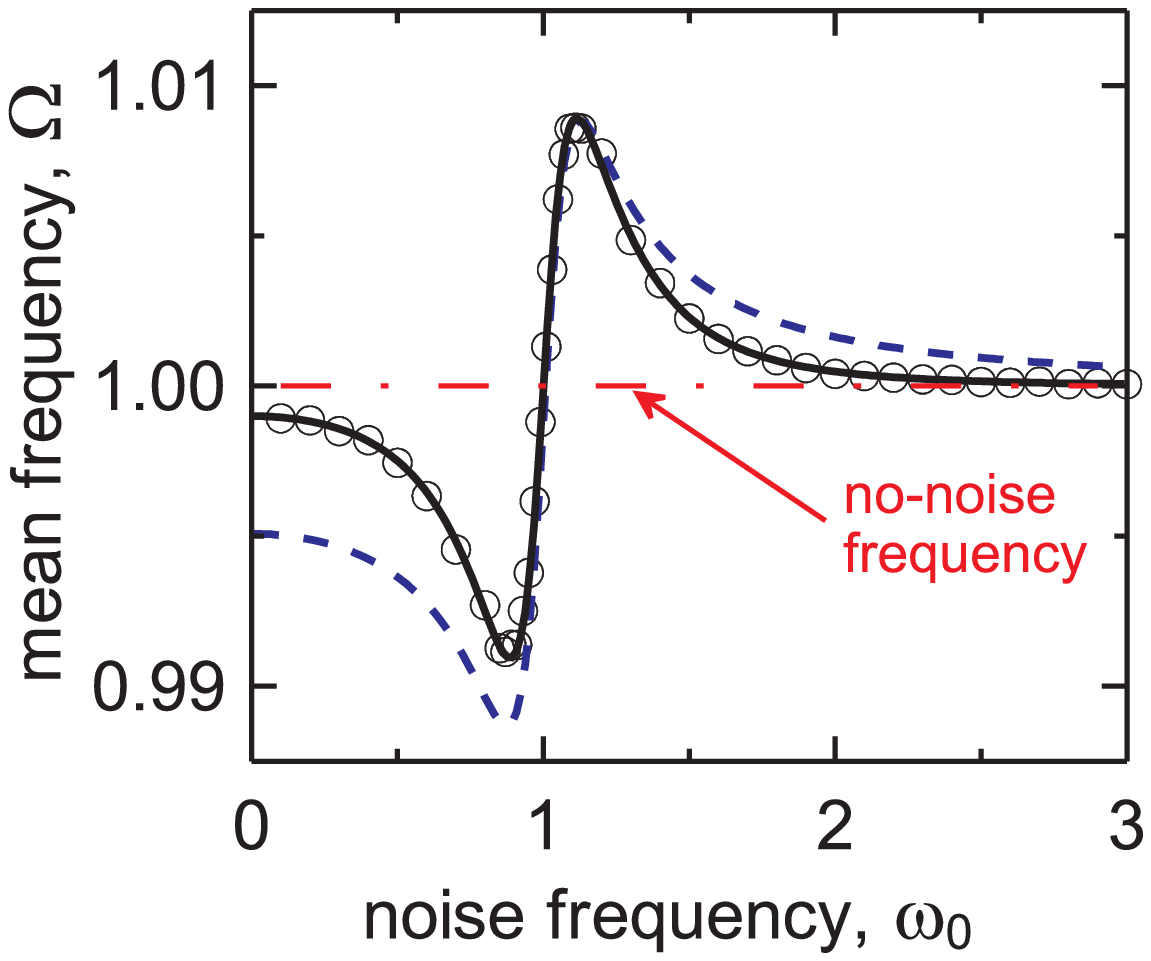}
 \quad
 (b)\hspace{-5mm}
 \includegraphics[width=0.222\textwidth]%
 {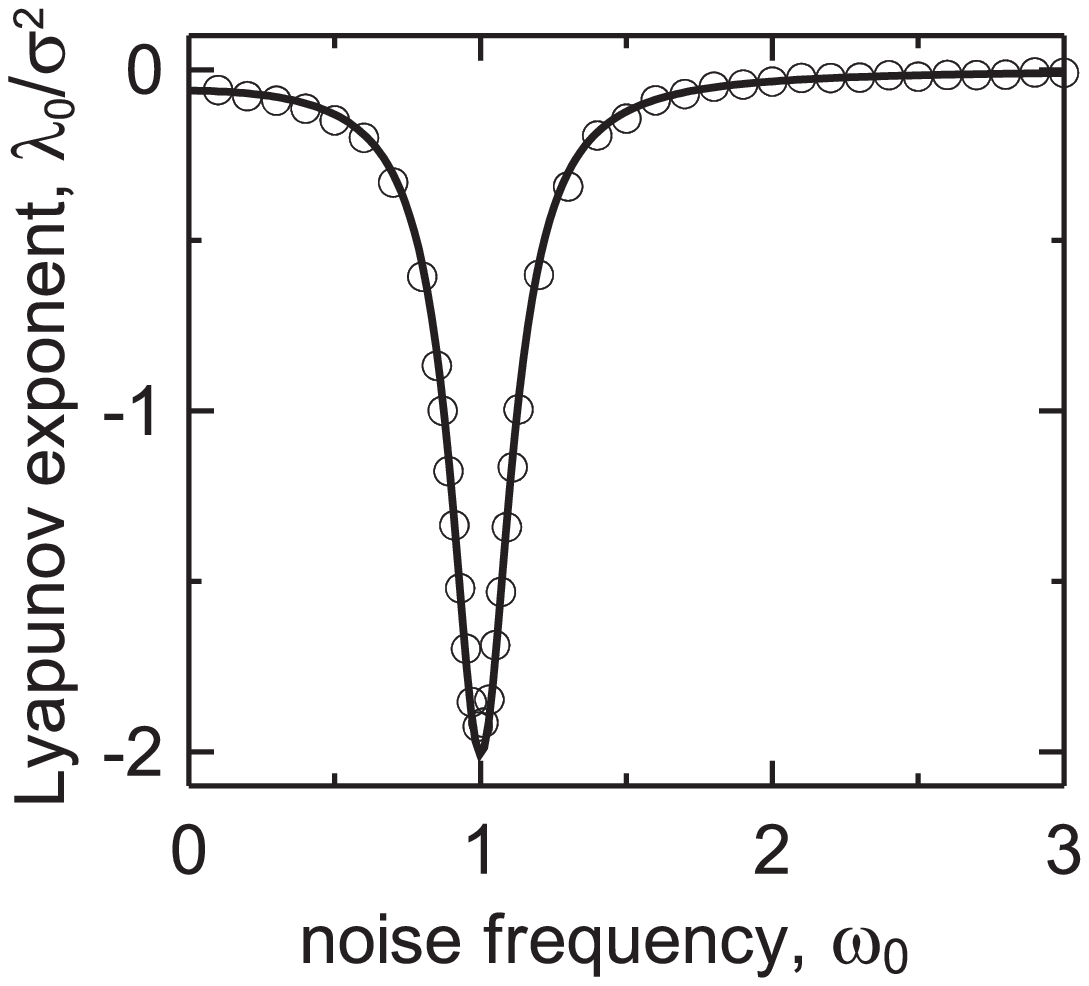}
}
  \caption{Hopf oscillator (\ref{eq-3-01}) subject to phase noise
$\eta(t)=\sqrt{2}\cos[\omega_0t+\sqrt{\gamma}\int^t\xi(t_1)dt_1]$
for $\sigma=0.1$, $\gamma=0.125$, $\lambda=0.4$. Circles:
numerical simulation, solid line: analytical theory
[Eqs.\,(\ref{eq-3-03}) and (\ref{eq-3-04})], dashed line:
analytical theory disregarding the amplitude degree of freedom.
 }
  \label{fig2}
\end{figure}

Another important particular opportunity yielded by the theory we
developed is the treatment of the effect of the phase noise,
$\eta(t)=\sqrt{2}\cos[\omega_0t+\sqrt{\gamma}\int^t\xi(t_1)dt_1]$.
With the noise autocorrelation function
$C(\tau)=\cos(\omega_0\tau)\exp(-\gamma|\tau|)$ one can evaluate
quantifiers, $\Omega$ and $\lambda_0$. In Fig.\,\ref{fig2} the
results of numerical simulation for the Hopf oscillator
[Eq.\,(\ref{eq-3-01})] subject to the phase noise are compared to
the analytical theory. Two points are worth emphasizing here:
(i)~Now we have the phase description for general oscillators
subject to noise which is the representative of signals of chaotic
and stochastic oscillators. This is important because it provides
us with a tool to analytically investigate the synchronizing
action of another oscillator (either chaotic or stochastic) on the
system under consideration in general. (ii)~The amplitude degree
of freedom is essential here: in the graph for the frequency
(Fig.\,\ref{fig2}), one can see how the analytical theory
neglecting the amplitude perturbations (dashed line) is far from
the real observations fairly fitted by the theory we have
developed. The most remarkable effects here are observed when the
characteristic noise correlation time $2\pi/\omega_0$ is
commensurable with the natural oscillation period of the system,
that is nonsmall, meanwhile the earlier studies were not able to
deal with such a case.

Summarizing, we have derived the reduced phase equation for limit
cycle oscillators subject to general non-Gaussian noise. The
derived phase equation correctly provides the mean frequency, the
phase diffusion constant and the Lyapunov exponent. Since the
noise-induced shift of the mean frequency means the shift of the
resonant frequency for entrainment by external
forcing~\cite{Yoshimura-Arai-2008,Yoshimura-2010}, our result for
mean frequency is immediately relevant for all investigations
concerning collective phenomena in networks of coupled
oscillators, {\it
e.g.},~\cite{Kuramoto-2003,collective_dynamics,neuronal_networks},
where noise is unavoidably present. In particular the theory is
valid for noise which is the representative of signals of chaotic
and stochastic oscillators and thus may provide an accurate
analytical tool to investigate their synchronizing action. For the
Lyapunov exponent, importance of the amplitude degrees of freedom
has been proven, though their average impact on the system
stability vanishes in the leading order of accuracy. This implies
that the analytical theories in earlier studies on the phase
diffusion and the Lyapunov exponent, where the amplitude degree of
freedom was disregarded ({\it e.g.},~\cite{Lyapunov_exponent}),
remain generally correct. The theory provides opportunity for
analytical investigation of the reliability of
neurons~\cite{Mainen-Sejnowski-1995} and consistency of
lasers~\cite{Uchida-Mcallister-Roy-2004} as well as the quality of
clocks, electric generators, lasers, {\it etc.} for general noise
and general limit cycle oscillators.

\acknowledgements{
D.S.G. acknowledges support from CRDF (Grant no.\ Y5--P--09--01).
J.-N.T acknowledges support from JST PRESTO and MEXT Japan (no.\
20700304). H.N. thanks MEXT, Japan (Grant no.\ 22684020). G.B.E.
acknowledges support from NSF DMS 0817131.




\newpage

\section{\sf \large Auxiliary material for Letter
``Dynamics of Limit Cycle Oscillator Subject to General Noise''}
{\center\sf\large
\noindent by Denis S.\ Goldobin, Jun-nosuke Teramae,\\
Hiroya Nakao, and G.\ Bard Ermentrout

}

\vspace{15pt}

Here we provide a rigorous derivation of Eqs.\,(1)--(2) of the
main article~\cite{Goldobin_etal-2010} (and their $N$-dimensional
version) governing evolution of a general dynamical system in the
basin of attraction of the limit cycle for the noise-free case.
The only restriction we impose is the differentiability of the
phase flux in the basin of attraction. This condition allows
employment of Taylor series and is fulfilled for a general system.

We would like to stress that Eqs.\,(1)--(2) are regarded as a
conventional paradigm for limit cycle systems, and
Ref.\,\cite{Yoshimura-2010} and this auxiliary materials serve the
purpose to confirm that this paradigm, intuitively adopted by
community, is an accurate description but not simply a model
catching key features of the oscillatory dynamics.

\section{2D phase space}
First let us consider two-dimensional case. We recall that the
phase can be introduced as the coordinate along the cycle so that,
it grows uniformly and increases by $2\pi$ for one revolution of
the system. The phase is governed by the equation
\begin{equation}
\dot{\phi}=\omega,
\label{aux01}
\end{equation}
where $\omega=2\pi/T$ is the cyclic frequency of oscillations, $T$
is the period. The phase can be extended to the whole basin of
attraction of the limit cycle so that Eq.\,(\ref{aux01}) holds
valid all over the basin~\cite{Guckenheimer-1975}. Let us briefly
outline the geometric explanation for this fact. We take the state
${\bf A}_0$ on the phase plane (see Fig.\ref{figaux01}) and let
the system evolve for the time period $T$, the new state is ${\bf
A}_1$. In its turn, ${\bf A}_1$ evolves for the same time period
to ${\bf A}_2$, and so forth. The sequence of ${\bf A}_n$ tends to
the limit cycle and ${\bf A}_\infty$ belongs to it. One can
connect points ${\bf A}_0$ and ${\bf A}_1$ by an arc, which can
deviate from the linear segment connecting these points. After
each iteration for one period $T$ the arc ${\bf A}_n{\bf A}_{n+1}$
turns into an arc, connecting points ${\bf A}_{n+1}$ and ${\bf
A}_{n+2}$. In such a way we end up with a curve running through
the points ${\bf A}_0$, ${\bf A}_1$, ${\bf A}_2$,..., ${\bf
A}_\infty$. This curve can be assigned the value of phase $\phi$
at point ${\bf A}_\infty$ of the limit cycle; in the literature,
it is referred to as {\em isochron}. Obviously, such definition of
phase $\phi$ is not unambiguous because there are infinitely many
arcs connecting ${\bf A}_0$ and ${\bf A}_1$; however it becomes
unambiguous when one claims the curve running through ${\bf A}_n$
to be smooth. Possibility to construct the field of phase $\phi$
all over the attraction basin is a well established fact and the
phase field were, for instance, numerically reconstructed for the
entire phase plane of the FitzHugh--Nagumo oscillator
in~\cite{Arai-Nakao-2008}.

\begin{figure}[!t]
\center{
 \includegraphics[width=0.35\textwidth]%
 {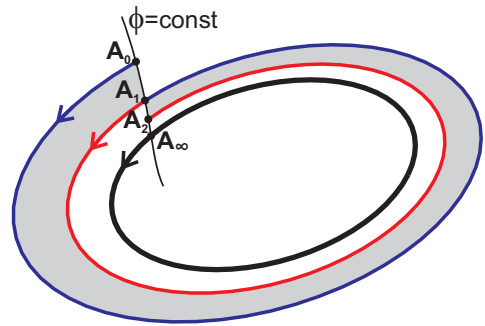}
}
  \caption{Sketch of construction of the field of $\phi$.
 }
  \label{figaux01}
\end{figure}

Now we have to complete construction of the coordinate grid with
the coordinate measuring the deviation from the limit cycle. It is
frequently referred to as the amplitude. However, one has to keep
in mind that it is rather deviation of the amplitude from the
value corresponding to the limit cycle, but not the conventional
amplitude. Let $\rho$ measures the length along the isochrones and
$\rho=0$ features the position on the limit cycle. Therefore,
\begin{eqnarray}
&&\dot{\phi}=\omega\,,
\label{aux02}\\
&&\dot{\rho}=F(\phi,\rho)=-\lambda(\phi)\,\rho+F_2(\phi,\rho)\,,
\label{aux03}
\end{eqnarray}
where $F_2(\phi,\rho)$ is a function which's Taylor series with
respect to $\rho$ starts with the term $\propto\rho^2$ or higher
powers.

We want to scale the variable $\rho$: we replace variable $\rho$
with $r=h(\phi,\rho)$ such that Eq.\,(\ref{aux03}) turns into
\begin{equation}
\dot{r}=-\mu r,
\label{aux04}
\end{equation}
where $\mu=(2\pi)^{-1}\int_0^{2\pi}\lambda(\phi)d\phi$ is the
average amplitude decay rate near the cycle. Now we have to
reconstruct the function $h(\phi,\rho)$ from
Eqs.\,(\ref{aux02})--(\ref{aux04}).

Let us consider isophase line $\phi=0$. The phase flow induces
mapping for $\rho$ on this line; the state $\rho(t=0)=\rho_0$ on
this line evolves after one revolution to
\[
\rho(t=2\pi/\omega)=G(\rho_0)=:
\sum_{n=1}^\infty G_n\,\rho_0^n\,;
\]
the function $G(\rho_0)$ has to be calculated from integration of
the equation system~(\ref{aux02}) and (\ref{aux03}). From
Eq.\,(\ref{aux04}),
\[
r(t=2\pi/\omega)=\Lambda\,r_0,
\]
where $\Lambda:=\exp(-2\pi\mu/\omega)$.

Matching the maps for $\rho$ and $r$, one can find the function
$h_0(\rho):=h(\phi=0,\rho)$. Indeed,
\[
r(2\pi/\omega)=h_0(\rho(2\pi/\omega))
=h_0(G(\rho_0))=h_0(G(h_0^{(-1)}(r_0))),
\]
where $h_0^{(-1)}$ is the inverse function of $h_0$. Function
$h_0(\rho)$ is a one-to-one (monotonously growing) function and,
therefore, $h_0^{(-1)}$ is well-defined. On the other hand,
$r(2\pi/\omega)=\Lambda\,r_0$. Equating two expressions for
$r(2\pi/\omega)$ and applying the function $h_0^{(-1)}$ to the
both sides of the equality, one finds
\begin{equation}
G(h_0^{(-1)}(r_0))=h_0^{(-1)}(\Lambda\,r_0).
\label{aux05}
\end{equation}
This equation can be resolved, {\it e.g.}, in terms of Taylor
series $G(r)=\sum_{n=1}^{\infty}G_n\,r^n$ and
$h_0^{(-1)}(r)=\sum_{n=1}^{\infty}a_n\,r^n$;
\begin{equation}
\sum_{n=1}^{\infty}G_n\bigg[
\sum_{m=1}^{\infty}a_m\,r_0^m\bigg]^n
=\sum_{n=1}^{\infty}a_n\Lambda^nr_0^n.
\label{aux06}
\end{equation}
Now one have to collect and equate terms with equal powers of
$r_0$. Thus,
\\
(1) for $r_0$:
\qquad $G_1a_1-a_1\Lambda=0$.
\\
here we found an obvious claim, $G_1=\Lambda$, which follows from
the fact that linearized in $\rho$ evolution of small deviations
from the limit cycle is determined by the multiplier $\Lambda$.
Coefficient $a_1$ remains undetermined because, in fact, no scale
for $r$ is imposed by our construction and we are free to choose
$a_1=1$.
\\
(2) for $r_0^2$:
\qquad $G_1a_2+G_2a_1^2-a_2\Lambda^2=0$.
\\
Hence,
\[
a_2=-\frac{G_2a_1^2}{G_1-\Lambda^2}
\stackrel{a_1=1}{=}-\frac{G_2}{(1-\Lambda)\Lambda}.
\]
(3) for $r_0^3$:
\[
a_3=-\frac{2G_2a_1a_2+G_3a_1^3}{(1-\Lambda^2)\Lambda}
\stackrel{a_1=1}{=}-\frac{2G_2a_2+G_3}{(1-\Lambda^2)\Lambda}.
\]
Hereby one can reconstruct the sequence of $a_n$ up to required
order of accuracy. With the function $h_0^{(-1)}(\rho)$ evaluated
one can find its inverse function $h_0(\rho)$.

Let us now consider Eq.\,(\ref{aux04}) with $r=h(\phi,\rho)$;
\[
\dot{r}(\phi,\rho)=\dot\phi\frac{\partial h}{\partial\phi}
+\dot\rho\frac{\partial h}{\partial\rho}=-\mu\,h(\phi,\rho).
\]
Substitution of $\dot\phi$ and $\dot\rho$ from Eqs.\,(\ref{aux02})
and (\ref{aux03}) yields
\[
\omega\frac{\partial h}{\partial\phi}
+F(\phi,\rho)\frac{\partial h}{\partial\rho}=-\mu\,h(\phi,\rho).
\]
One can consider this as an evolution equation
\begin{equation}
\frac{\partial h}{\partial\phi}=-\frac{\mu}{\omega}h(\phi,\rho)
-\frac{F(\phi,\rho)}{\omega}\frac{\partial h}{\partial\rho}
\label{aux07}
\end{equation}
with initial condition
\[
h(\phi=0,\rho)=h_0(\rho)
\]
calculated from Eq.\,(\ref{aux05}) or (\ref{aux06}) as described.

For small deviations $\rho$ (or $r$), when one can neglect
nonlinear terms in $F(\rho)$ and $G(\rho)$, one finds
$h_0(\rho)=\rho$ and solution to Eq.\,(\ref{aux07}):
\begin{equation}
r=h(\phi,\rho)
 =\rho\exp\bigg[\frac{1}{\omega}
 \int\limits_0^\phi(\lambda(\psi)-\mu)d\psi\bigg]+O(\rho^2).
\label{aux08}
\end{equation}

The particular result for small deviations, Eq.\,(\ref{aux08}),
can be found in~\cite{Yoshimura-2010}. This result means that with
an appropriate choice of the coordinates one can obtain a constant
decay rate for amplitude deviations even when in ordinary
coordinates one can observe positive instantaneous Lyapunov
exponent $(-\lambda(\phi))$ meaning local divergence of
trajectories. However, here we have shown the regular procedure
for constructing parameterization $(\phi,r)$ such that, the
evolution of two-dimensional dynamical system is accurately
described by equations
\[
\dot{\phi}=\omega,\qquad\qquad
\dot{r}=-\mu\,r,
\]
not only for small deviations, but all over the attraction basin
of the limit cycle. In the presence of noise $\sigma\eta(t)$ one
finds
\begin{eqnarray}
 &&\dot{\phi}=\omega+\sigma\,Z_\phi(\phi,r)\,\eta(t)\,,
\\
 &&\dot{r}=-\mu\,r+\sigma\,Z_r(\phi,r)\,\eta(t)\,,
\end{eqnarray}
where $Z_\phi$ and $Z_r$ are sensitivity functions. This is the
equation system (1)--(2) of the main
article~\cite{Goldobin_etal-2010} up to notations.

\section{$N$-D phase space}
In higher dimensions we restrict our consideration to the case of
small deviations because the procedure for consideration of
nonlinearities is principally the same as for the two-dimensional
case, but significantly more lengthy. The deviation from the limit
cycle is now parameterized by $(N-1)$-dimensional vector
$\vec{\rho}$. For linearized case, one can choose the point
$\phi=0$ and construct the linear mapping ${\bf A}$;
\[
\vec{\rho}(\phi=0)={\bf A}\cdot\vec{\rho}(\phi=1)\,.
\]
As long as matrix ${\bf A}$ possesses only positive or complex
eigenvalues (multipliers) $\Lambda_j=\exp(2\pi\mu_j/\omega)$ with
eigenvectors $\vec{\rho}_j$, one can choose the coordinate grid
 $(\phi,\vec{r})$ such that
 $\vec{\rho}(t)=\sum_jr_j(t)\vec{\rho}_j(\phi)e^{-\mu_j\phi}$,
where
 $\vec\rho_j(\phi)=\vec\rho(t=\phi)|_{\vec\rho(t=0)=\vec\rho_j}$.
Then
\[
\dot{r_j}=-\mu_j\,r_j+O(r^2)\,.
\]
This equation system was assumed for derivation of the phase
reduction equation (9) and quantifiers of the dynamics (10) and
(11) in the main paper.

In fact, our constructions correspond to the employment of the
basis of Floquet eigenvectors and development of this methodology
for the case of nonlinear equations.

\end{document}